\documentclass[runningheads]{llncs}
\usepackage[T1]{fontenc}
\usepackage{graphicx}
\usepackage{graphicx}
\usepackage{booktabs}
\usepackage{amssymb}
\usepackage{amsmath}
\usepackage{color}
\usepackage{soul}
\usepackage{enumitem}
\usepackage{lineno,hyperref}
\usepackage{colortbl}
\usepackage{graphicx}
\usepackage{fontawesome}
\definecolor{mygray}{gray}{.9}
\let\OLDthebibliography\thebibliography
\renewcommand\thebibliography[1]{
  \OLDthebibliography{#1}
  \setlength{\parskip}{0pt}
  \setlength{\itemsep}{0pt plus 0.3ex}
}

\begin{document}
\title{Expert Insight-Enhanced Follow-up Chest X-Ray Summary Generation}
%
%
\author{Zhichuan Wang\inst{1,2}\orcidID{0009-0003-0373-3971} \and Kinhei Lee
\inst{1}\orcidID{0009-0000-6832-9863} \and 
Qiao Deng\inst{1,2}\orcidID{0009-0004-6233-9840} 
\and 
 Tiffany Y. So\inst{1}\orcidID{0000-0001-8268-0721} 
\and
Wan Hang Chiu\inst{3}\orcidID{0000-0002-7930-1193}
\and
Yeung Yu Hui\inst{5}\orcidID{0000-0002-4887-7974}
\and
Bingjing Zhou\inst{1}\orcidID{0009-0007-9734-4721}
\and
Edward S. Hui\inst{1,2,4}\textsuperscript{\faEnvelope} \orcidID{0000-0002-1761-0169}
}
\authorrunning{Z. Wang et al.}
%
\institute{Department of Imaging and Interventional Radiology, the Chinese University of Hong Kong, Hong Kong, China \and
CU Lab for AI in Radiology (CLAIR), The Chinese University of Hong Kong, HKSAR, China
\and
Hospital Authority, Hong Kong, China \and
Department of Psychiatry, the Chinese University of Hong Kong, China \\
\and
China Unicom Global Limited \\
\email{edward.s.hui@gmail.com}\\
}
%
\maketitle              
\begin{abstract}
A chest X-ray radiology report describes abnormal findings not only from X-ray obtained at current examination, but also findings on disease progression or change in device placement with reference to the X-ray from previous examination. Majority of the efforts on automatic generation of radiology report pertain to reporting the former, but not the latter, type of findings. To the best of the authors' knowledge, there is only one work dedicated to generating summary of the latter findings, i.e., follow-up summary. In this study, we therefore propose a transformer-based framework to tackle this task. Motivated by our observations on the significance of medical lexicon on the fidelity of summary generation, we introduce two mechanisms to bestow expert insight to our model, namely expert soft guidance and masked entity modeling loss. The former mechanism employs a pretrained expert disease classifier to guide the presence level of specific abnormalities, while the latter directs the model's attention toward medical lexicon. Extensive experiments were conducted to demonstrate that the performance of our model is competitive with or exceeds the state-of-the-art.

\end{abstract}
\section{Introduction}
\begin{figure}[!t]
\centering
\includegraphics[width=0.8\textwidth]{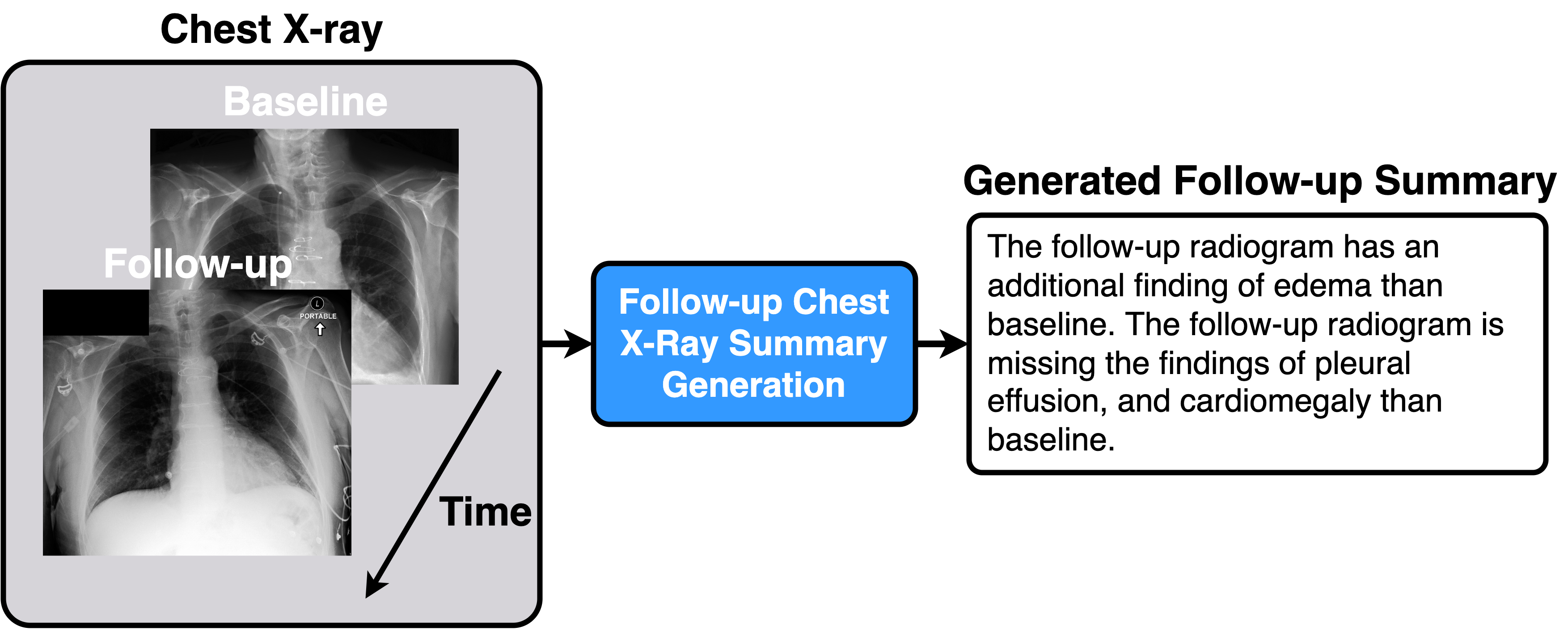}
\caption{A schematic of our proposed model that generates a textural summary of disease progression from a pair of chest X-rays taken at follow-up and baseline examinations.}
\label{illustration}
\end{figure}

Chest X-ray radiology report generation can automatically report radiological findings on radiogram in an end-to-end manner \cite{tanida2023interactive,miura2020improving}, potentially providing a means to alleviate the workloads of radiologists. A major shortcoming of the majority of these models is that
they are limited to reporting abnormalities in a given chest X-ray examination. Typically, radiologist writes reports not only based on a single (follow-up) X-ray examination but also by incorporating information from the X-ray from previous check-up to discern disease progression and/or change in device placement. Automatic generation of this type of report is known as follow-up report generation, a clinically critical yet largely understudied area. To the best of the knowledge, there is only one work \cite{hu2023expert} focusing on this field. We therefore aim to develop an end-to-end deep learning model to generate textural summary of disease progression from chest X-ray obtained at follow-up and baseline examinations (see Figure \ref{illustration}). 

However, majority of the current models, namely follow-up summary generation \cite{hu2023expert}, single radiogram report generation \cite{tanida2023interactive,miura2020improving}, or natural image difference caption \cite{qiu2021describing,yao2022image}, overlook the importance of meaningful entities, \textit{i.e.}, the medical lexicon within the domain of follow-up summary generation. Specifically, these models typically generate a sentence word by word under the supervision of the masked language modeling loss. In this paradigm, these models pay equal attention to each word, which may not be optimal, as words related to abnormalities should deserve more attention than other words. 
This asymmetry results in a relatively lower accuracy in predicting abnormality-related words, thereby necessitating additional mechanism to improve the fidelity of follow-up summary generation.

Toward this end, we propose two mechanisms to bestow expert insight on follow-up chest X-ray summary generation, namely expert soft guidance and masked entity modeling loss. The former employs a pretrained expert disease classifier to provide complementary guidance on the presence level of certain diseases, while the latter directs the model's attention toward abnormality-related words. Our contributions are:  
\begin{itemize}[noitemsep] 
  \item A framework is proposed to address the follow-up chest X-ray summary generation task, a clinically critical but largely understudied area. 
  \item Two mechanisms, namely expert soft guidance and masked entity modeling loss, are proposed to bestow expert insight on disease-related words.
  \item Comprehensive experiments were conducted to demonstrate that the performance of our model is competitive with or exceeds the state-of-the-art.
\end{itemize}




\section{Related Work}
The work of Hu et al. \cite{hu2023expert} is arguably the inaugural work to tackle the task of follow-up summary generation. The authors have assembled a dataset, dubbed MIMIC-Diff-VQA, of radiogram pairs and their corresponding follow-up chest X-ray summaries using the MIMIC-CXR dataset \cite{johnson2019mimic}. They proposed EKAID, a graph-based network, to extract relationships between anatomical regions and diseases, with commendable performance. 


The most related field in natural images is image difference caption, which aims to analyze the semantic differences in a given pair of images. The Spot-the-diff dataset was collected by \cite{jhamtani2018learning} in 2018, marking the introduction of the image difference caption task. Later, Qiu et al. introduced a transformer-based model, MCCFormers, to capture the intra and inter-image relations, further improving the performance \cite{qiu2021describing}. With the pretraining paradigm demonstrating a powerful ability in vision-language tasks, Yao et al. devised three pretraining tasks, IDCPCL, to align visual and textural features, achieving superior results compared to previous works \cite{yao2022image}.


\begin{figure*}[ht]
\centering
\includegraphics[width=1.0\textwidth]{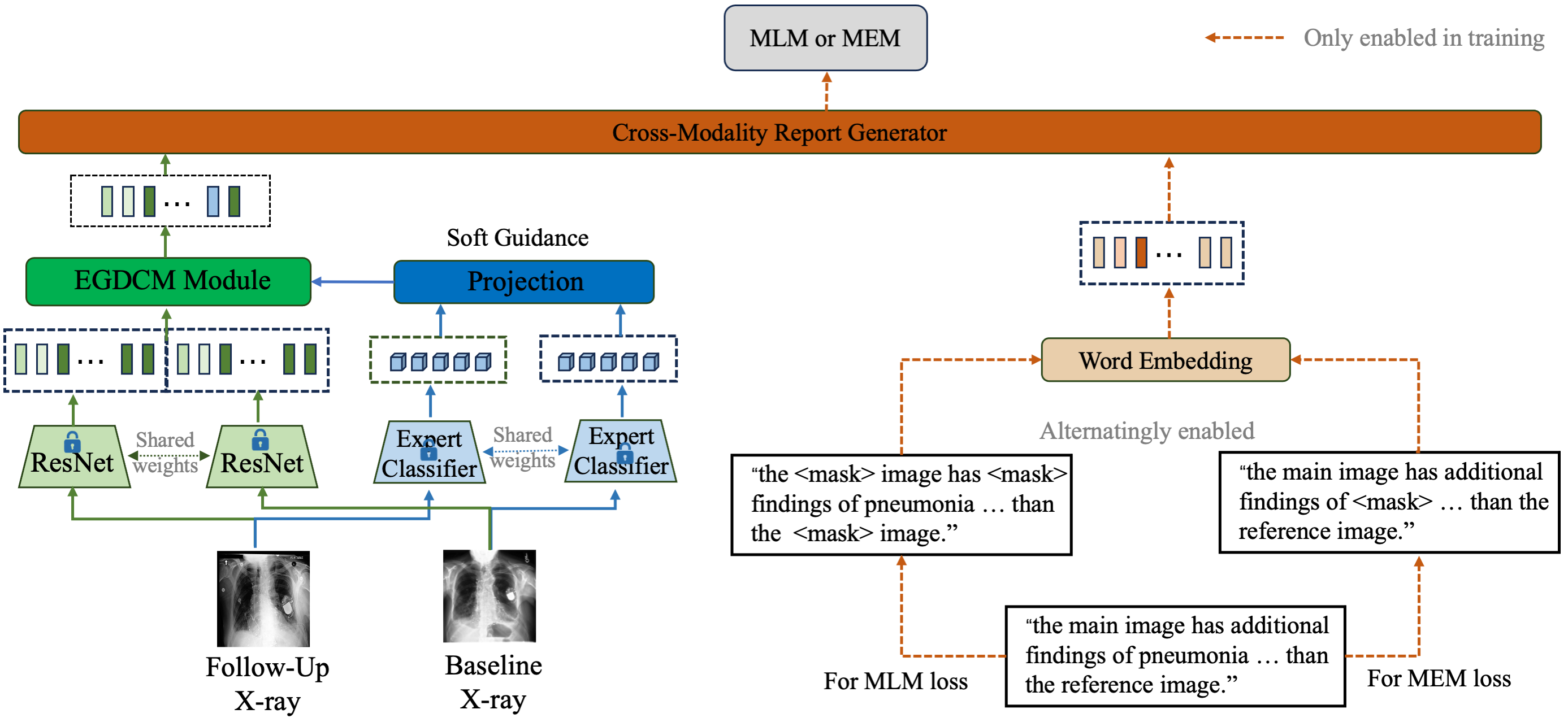}
\caption{Illustration of our proposed framework, expert insight-enhanced (EIE) follow-up chest X-ray summary generation. The dotted arrows indicate the branches that are only enabled in the training phase, and the symbol of lock indicates module that are frozen. EGDCM: expert guided difference capture module; MLM: masked language modelling; MEM: masked entity modelling.}
\label{framework}
\end{figure*}

\section{Method}
\subsection{Preliminaries}
Given an X-ray pair $(\mathbf{I}^f,\mathbf{I}^b)$ from the same patient, where $\mathbf{I}^f$ and $\mathbf{I}^b$ respectively represent X-ray images obtained at follow-up and baseline examinations. Our goal is to generate a summary of radiological findings ${\mathbf{s}}$ that describes disease progression and/or change in device placement at the time of follow-up X-ray examination. Such summary is dubbed as follow-up summary from hereon.

During the training phase, X-ray pair $\{({\mathbf{I}^f}^{(i)},{\mathbf{I}^b}^{(i)})\}_{i\in\mathbb{D}}$ and the corresponding ground-truth follow-up summary $\{\mathbf{s}^{(i)}\}_{i\in\mathbb{D}}$ from training dataset $\mathbb{D}$ were used to train the model. Building upon prior research \cite{yao2022image,qiu2021describing}, every word in $\mathbf{s}$ underwent tokenization and was subsequently mapped to word embeddings via an embedding layer initiated with training from the ground up. Three special tokens were added to facilitate the training process, namely $\texttt{[CLS]}$, $\texttt{[BOS]}$ and $\texttt{[EOS]}$. $\texttt{[CLS]}$ was added to merge the information of all word tokens, and $\texttt{[BOS]}$ and $\texttt{[EOS]}$ were used to identify the beginning or ending of a sentence. The text input can be represented as:
\begin{equation}
\bar{\mathbf{s}} = \left(\texttt{[CLS]}, \texttt{[BOS]}, w_1,\dots,w_{N_s}, \texttt{[EOS]} \right),
\end{equation}
where $w_i$ represents the token of the $i^{th}$ word in $\mathbf{s}$. Similar to the approach in \cite{yao2022image}, a pretrained ResNet101 \cite{he2016deep} was adopted to obtain the image features from the X-ray pair, which were then flattened into a sequence of image tokens:
\begin{align}
  \mathbf{V}^b = \left(\texttt{[Xray1]}, \mathbf{V}_0^b, \dots, \mathbf{V}_{N_v}^b\right), \quad \mathbf{V}^f = \left(\texttt{[Xray2]}, \mathbf{V}_0^f,\dots, \mathbf{V}_{N_v}^f\right), 
\end{align}
where $\texttt{[Xray1]}$ and $\texttt{[Xray2]}$ are two tokens for capturing the comprehensive representations of individual X-ray images, and $\mathbf{V}_i^b$ and $\mathbf{V}_i^f$ represent the $i^{th}$ image token of the baseline and follow-up X-ray, respectively. Furthermore, fixed positional embeddings and type embeddings were incorporated, with the former designed to denote token positions and the latter employed to distinguish the source of a particular token, such as $\mathbf{V}^f$, $\mathbf{V}^b$, or $\bar{\mathbf{s}}$.

\subsection{Model Architecture}
The overall architecture of our proposed model, expert insight-enhanced (EIE) follow-up chest X-ray summary generation, is illustrated in Figure~\ref{framework}. A pair of X-ray images are fed into a pretrained ResNet to extract image features, serving as image tokens after flattening. The same pair of X-ray are also input into a pretrained expert disease classifier to obtain the probability of the presence of certain diseases, which is then projected into the same dimension as image tokens. Image and soft-guidance tokens are fed into the transformer-based expert guided difference capture module (EGDCM) to obtain the difference representation. The output tokens of EGDCM and word tokens are input into the cross-modality summary generator to produce the follow-up chest X-ray summary. The generation is supervised by the masked language modeling (MLM) loss and masked entity modeling (MEM) loss. 
\subsubsection{Expert Soft Guidance}
To bestow expert insight to the model for enhancing the prediction of abnormality-related words, known as entities, complementary guidance on the presence level of certain diseases was provided by a pretrained multi-class expert classifier.

Given an X-ray pair $(\mathbf{I}^f, \mathbf{I}^b)$, a pretrained multi-class disease classifier was employed to generate the probability of the presence of a number of abnormalities, denoted as $(\mathbf{p}^f, \mathbf{p}^b)$. Rather than using a hard threshold to binarize these probabilities, they were directly mapped into the same feature space of image tokens through a projection module. Subsequently, the guidance tokens were obtained and used to guide the expert guided difference capture module (EGDCM) in capturing a more disease-related representation. Formally, the process of obtaining the soft guidance can be denoted as:
\begin{align}
  \mathbf{p}^j &= \text{F}_e(\mathbf{I}^j), \\
 \mathbf{g}^j &= \text{Proj} (\mathbf{p}^j) \quad \text{for } j \in \{f,b\},
\end{align}
where $\text{F}_e$ denotes the expert classifier, $\text{Proj}(\cdot)$ denotes the projection layer, and $\mathbf{g}^j$ represents the soft-guidance token. Similar to image tokens, the soft-guidance tokens $\mathbf{g}^j$ are appended with fixed positional embeddings and type embeddings.

\subsubsection{Expert-Guided Difference Capture Module} 
We propose an expert-guided difference capture module (EGDCM) $\text{F}_{\text{diff}}$ to extract the representations of the difference in the image features of two different examinations. To better model long-distance dependencies and merge information from image and texts, EGDCM is composed of a two-layer transformer, which takes all image tokens and guidance tokens as input. Formally, the process can be written as:
\begin{align}
  \mathbf{V}^b_E = \left(\texttt{[Xray1]}, \mathbf{V}_0^b, \dots, \mathbf{V}_{N_v}^b, \mathbf{g}^b\right) &, \quad \mathbf{V}^f_E = \left(\texttt{[Xray2]}, \mathbf{V}_0^f,\dots, \mathbf{V}_{N_v}^f, \mathbf{g}^f\right) \\
\{\bar{\mathbf{V}}^f,\bar{\mathbf{V}}^b\} & = \text{F}_{\text{diff}} (\{\mathbf{V}^f_E,\mathbf{V}^b_E\}).
\end{align}

\subsubsection{Cross-Modality Follow-Up Summary Generator}
A three-layer transformer was employed for the cross-modality report generator $\text{F}_{\text{gen}}$, which took the image difference tokens and summary tokens as inputs:
\begin{equation}
(\hat{\mathbf{V}}^f,\hat{\mathbf{V}}^b,\hat{\mathbf{s}}) = \text{F}_{\text{gen}}(\bar{\mathbf{V}}^f, \bar{\mathbf{V}}^b, \bar{\mathbf{s}})
\end{equation}

\subsection{Loss Functions}
To endow the model with the ability to generate follow-up chest X-ray summary, the widely used MLM loss \cite{yao2022image,devlin2018bert} was adopted. Furthermore, we propose a new MEM loss, an extension of MLM, to enhance the prediction of the abnormality-related words. 

\subsubsection{Masked Language Modeling Loss}
In contrast to many vision-language pretraining works, the uni-directional MLM loss was employed to emulate the sentence generation process of inference. In other words, instead of considering all text tokens, the prediction of the $i^{th}$ (masked) word was based solely on its preceding words $w_{< i}$ and other image-related tokens $(\hat{\mathbf{V}}^f, \hat{\mathbf{V}}^b)$. The same masking strategy as BERT \cite{devlin2018bert} was used, where 15\% of the input text tokens were randomly selected, with 80\% of which masked by the special token $\texttt{[MASK]}$, 10\% substituted with words randomly chosen in vocabulary, and the remaining 10\% left unchanged. To obtain the final predictions, the masked tokens were input into a linear classifier. Formally, the uni-directional MLM loss over $\mathbb{D}$ can be written as:
\begin{align}
L_{\text{MLM}} = \mathbb{E}_{\mathbb{D}}[-\log P_{\theta}(w_i & | w_{<i}, \hat{\mathbf{V}}^f, \hat{\mathbf{V}}^b)] \quad \text{for } \forall i \in \{1,\cdots,N_s \},
\end{align}
where $\theta$ is the trainable parameters of the model.

\begin{table*}[!t]
\renewcommand{\arraystretch}{1.}

\centering

\resizebox{1.\linewidth}{!}{
\begin{tabular}{cccccccccc}
\toprule
Method             &BLEU-1          &BLEU-2        &BLEU-3          &BLEU-4          &METEOR           &ROUGE$_\text{L}$           &CIDEr   &Acc5      &Acc14 \\
\hline
\midrule
$\text{MCCFormers}^{\dagger}$         &0.214           &0.190         &0.170           &0.153           &0.319            &0.340           &0.000       &-       &-\\
$\text{IDCPCL}^{\dagger}$	        &0.614           &0.541         &0.474           &0.414           &0.303             &0.582          &0.703         &-       &-\\
EKAID              &0.628		    &0.553         &0.491           &0.434           &0.339            &0.577           &1.027           &$\text{68.21}^{\ddagger}$     &$\text{81.00}^{\ddagger}$\\
\hline
\hline
EIE-base          &0.574           &0.510         &0.456           &0.407           &0.416            &0.590           &0.968       &59.94    &77.57\\
EIE-mem           &0.614		    &0.547         &0.492           &0.441           &0.405            &0.616           &1.421       &65.89       &79.14\\
EIE-esg           &0.638           &0.573         &0.519           &0.469           &0.399            &0.636           &1.615         &68.33       &81.63 \\
\rowcolor{mygray} 
EIE-all               &0.646           &0.583         &0.528           &0.477           &0.401            &0.635           &1.698     &70.98   &81.47\\

\bottomrule

\end{tabular}
}
\caption{Comparison with State-of-the-Art (SOTA): \textbf{EIE-all} surpasses existing SOTA by a substantial margin across all metrics. Furthermore, each proposed mechanism notably enhances the performance of the base model. The combination of both mechanisms leads to further improvement. $\dagger$ Results were reported from EKAID. $\ddagger$ Results were reproduced using the official codes of EKAID.}

\label{SOTA comparison}

\end{table*}

\subsubsection{Masked Entity Modeling Loss}
Prediction of entities, can be enhanced by directing model's attention toward these words through our proposed MEM. Masked entity modeling loss is formulated similar to the MLM loss, but with the masking of the selected entity words: 
\begin{equation}
L_{\text{MEM}} = \mathbb{E}_{D}[-\log P_{\theta}(w_i | w_{<i}, \hat{\mathbf{V}}^f, \hat{\mathbf{V}}^b)] \quad \text{for } w_i \in \mathcal{E}, 
\end{equation}
where $\mathcal{E}$ is the selected entity set. 

Given that expert soft guidance can only offer soft guidance for certain abnormalities, the MEM loss serves as a complement to expert soft guidance by enhancing the prediction of the remaining common and crucial abnormality-related words. For more details on entity selection, please refer to the appendix. 

It is worth noting that, thanks to the parallel computation of transformer blocks, the soft guidance tokens $\mathbf{g}^f$ and $\mathbf{g}^b$ can be easily injected into or removed from the difference capturing process. When the model was trained solely with the MLM loss and without enabling expert soft guidance, we name the model as \textbf{EIE-base}. When \textbf{EIE-base} was combined with expert soft guidance or MEM loss, we refer to it as \textbf{EIE-esg} and \textbf{EIE-mem}, respectively. When \textbf{EIE-base} was combined with both proposed mechanisms, it is denoted as \textbf{EIE-all}.




\subsection{Training Strategy}
When both of the MLM and MEM losses were enabled, an alternating optimization strategy was adopted. Specifically, an enabling probability $\alpha$ was assigned to the MEM loss. In a given training iteration, a random sample $r$ was drawn from a uniform distribution between 0 and 1. The MEM loss was employed if $r > \alpha$; the MLM loss was enabled otherwise. The optimization direction was controlled by adjusting the hyperparameter $\alpha$.

\begin{figure}[h]
    \centering
    \includegraphics[width=0.4\linewidth]{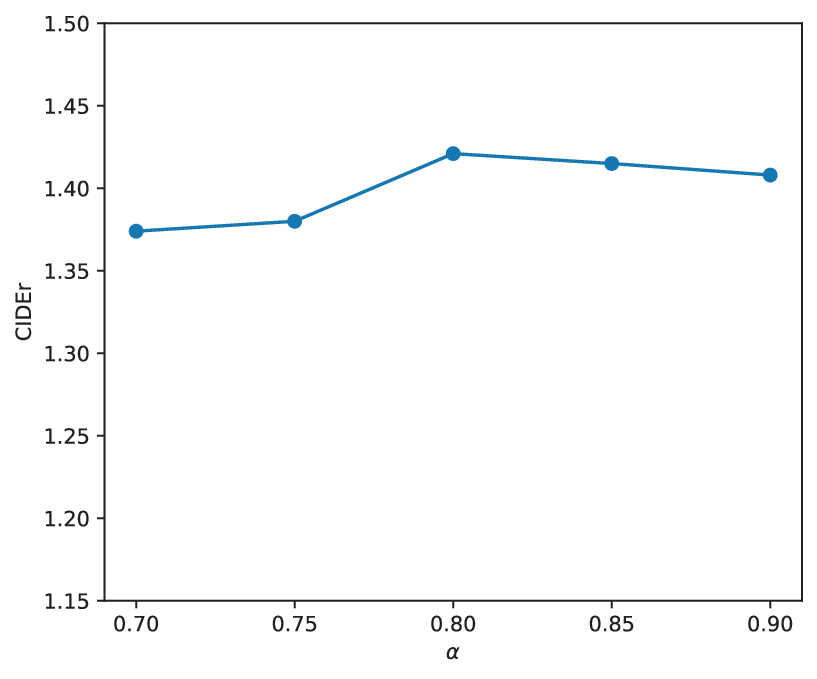}
    \caption{Hyperparameter $\alpha$ sensitivity study of EIE-mem.}
    \label{Hyper-parameter sensitive study}
\end{figure}

\section{Experiments}
\subsection{Dataset and Metrics}
Our model was evaluated on the MIMIC-Diff-VQA \cite{hu2023expert} dataset, which contains 164,324 X-ray pairs and the corresponding follow-up reports. Following \cite{hu2023expert,yao2022image}, widely used natural language generation metrics including BLEU \cite{papineni2002bleu}, METEOR \cite{denkowski2014meteor}, ROUGE$_{\text{L}}$ \cite{rouge2004package} and CIDEr \cite{vedantam2015cider} were adopted to evaluate the model on summary generation. Following \cite{miura2020improving}, our method was also evaluated using accuracy Acc5 and Acc14 to demonstrate its ability on abnormality recognition, where Acc5 denotes the micro average accuracy over 5 most common abnormalities and Acc14 is that over 14 abnormalities of CheXpert \cite{irvin2019chexpert}. Please refer to the appendix for more details on the abnormalities, the metrics and their computation. 


\subsection{Implementation Details}
Consistent with previous works \cite{yao2022image}, a pretrained ResNet101 was employed to extract grid features with a shape of (7, 7, 1024), which were then flattened into a sequence of image tokens with a shape of (49, 1024). For all transformer-based modules, the hidden feature size was 512, the number of multi-heads was 8, and the layer numbers were 2 and 3 for the EGDCM and cross-modality follow-up summary generator, respectively. The word embedding layers were learnt from scratch, and the feature dimension was set to 512. The hyperparameter $\alpha$ in the alternating optimization strategy was set to 0.8. The Adam optimizer was used with a learning rate of 3e-5, and the total training iteration was set to 100k. Our expert classifier was based on PCAM \cite{ye2020weakly}. For more details, please refer to the appendix. 

\subsection{Comparison with the State-of-the-Art}
As depicted in Table \ref{SOTA comparison}, with the exception of the \textbf{EIE-base} model, other versions of the proposed methods outperformed existing SOTA by a considerable margin, with \textbf{EIE-all} achieving the best performance. Taking \textbf{EIE-all} as an example, there were improvements of 1.8\%, 3.0\%, 3.7\%, and 4.3\% on the BLEU metrics. For METEOR, a performance gain up to 6.2\% was achieved. The increase was more pronounced on ROUGE$_{\text{L}}$ and CIDEr, with \textbf{EIE-all} boosting the performance by 5.8\% and 0.671, respectively. The relative increase in CIDEr was particularly large, at 65.3\% compared with EKAID. For Acc5 and Acc14, a 2.77\% and 0.47\% accuracy performance gain was obtained. These results demonstrate that our method performs better on both the report level and the clinical level. 

\subsection{Ablation Study}
As illustrated in Table \ref{SOTA comparison}, each proposed mechanism notably enhanced the performance of our base model, both at the generation and clinical levels. \textbf{EIE-mem} achieved improvements of 4.0\%, 3.7\%, 3.6\%, and 3.4\% on the BLEU metric, with a 2.6\% and 0.453 increase in ROUGE$_{\text{L}}$ and CIDEr, respectively. There were improvements in both Acc5 and Acc14, at 5.95\% and 1.57\% respectively. For \textbf{EIE-esg}, the performance gains were 6.4\%, 6.3\%, 6.3\%, and 6.2\% on BLEU, 8.39\% on Acc5, and 4.06\% on Acc14. ROUGE$_{\text{L}}$ increased from 59.0\% to 63.6\%, and CIDEr increased from 0.968 to 1.615. Notably, the relative increase in CIDEr was 66.84\%. Combining the MEM loss and expert soft guidance further improved performance, with \textbf{EIE-all} achieving gains of 7.2\%, 7.3\%, 7.2\%, 7.0\%, 4.5\%, 0.73, 11.04\%, and 3.9\% on BLEU, ROUGE$_{\text{L}}$, CIDEr, Acc5, and Acc14 respectively.

\begin{table}[t]
\renewcommand{\arraystretch}{1.}

\centering
\resizebox{0.9\linewidth}{!}{
\begin{tabular}{cccccccc}
\toprule
Threshold    
&BLEU1    &BLEU2    &BLEU3    &BLEU4    &METEOR    &ROUGE$_{\text{L}}$    &CIDEr  \\    
\midrule
0.4      
&0.548    &0.491    &0.445    &0.403    &0.304     &0.511    &1.557      \\
0.5                       
&0.578    &0.517    &0.470    &0.426    &0.317     &0.523    &1.579      \\
0.6       
&0.571    &0.511    &0.464    &0.421    &0.314     &0.520    &1.576     \\
\rowcolor{mygray} 
Soft guidance                      
&0.638    &0.573    &0.519    &0.469    &0.399     &0.636    &1.615    \\
\bottomrule
\end{tabular}
}
\caption{Comparison with hard versus soft guidance.}
\label{threshold table}
\end{table}
\subsection{Hyper-Parameter Sensitivity Study}
Several experiments were conducted to analyze how the performance of \textbf{EIE-mem} varied with the hyper-parameter $\alpha$, which controls the enable probability of the MEM loss. The results are presented in Figure \ref{Hyper-parameter sensitive study}. The CIDEr score ranged from 1.37 to 1.42, and surpassed 1.027, which was that of the existing state-of-the-art\cite{hu2023expert}. The best result of 1.42 was obtained when $\alpha = 0.8$. Consequently, we set $\alpha$ to 0.8 in all experiments.

\subsection{Comparison with Hard Guidance}
Several experiments were conducted to compare the performance of hard guidance with that of soft guidance. To obtain hard guidance, a sigmoid function was applied to the output of the expert classifier,  which was subsequently binarized using a threshold. The results are presented in Table \ref{threshold table}. Soft guidance outperformed hard guidance, particularly in METEOR and ROUGE$_{\text{L}}$ metrics. Setting the threshold at 0.5 yields the best results among all hard guidance experiments. Soft guidance achieved 6.0\%, 5.6\%, 4.9\%, and 4.3\% improvement in BLEU metric, 8.2\% in METEOR, 11.3\% in ROUGE$_{\text{L}}$, and 0.036 in CIDEr. The relative performance increases in METEOR and ROUGE$_{\text{L}}$ were 25.87\% and 21.6\%, respectively. These experiments affirm that soft guidance is more effective and robust, as it eliminates the need for careful hyperparameter selection.

\subsection{Lightweight EIE-all and more experiments}
In addition to the aforementioned experiments, the appendix contains supplementary experimental results and showcases, including a lightweight version of \textbf{EIE-all}. This variant utilized expert guidance during the training phase to enhance the knowledge learning, whilst omitting the expert soft guidance branch during inference. In addition, the performance of a model with a stronger and recently proposed expert classifier was also investigated, indicating that model performance can be further improved with better expert guidance. For more details, please refer to the appendix.

\section{Limitations}
Albeit exceeding the performance of SOTA, our proposed model warrants further improvement in its clinical accuracy. In addition, we will bestow our model the ability to predict the location of abnormalities in future studies.

\section{Conclusion}
A transformer-based framework was proposed to address the clinically critical yet largely understudied task of follow-up chest X-ray summary generation. Two mechanisms, namely expert soft guidance and masked entity modeling loss, were proposed to bestow expert insight on abnormality recognition for improving the overall model performance. Extensive experiments demonstrated that the performance of our model is competitive with or exceeds the state-of-the-art.





%

%
%
%
%

\bibliographystyle{splncs04}
\bibliography{icme2023template}

\end{document}


\def\x{{\mathbf x}}
\def\L{{\cal L}}

\title{Appendix}
%

\author{}
\institute{}
\maketitle

\section{Lightweight EIE-all}
\begin{figure}
    \centering
    \includegraphics[width=0.7\linewidth]{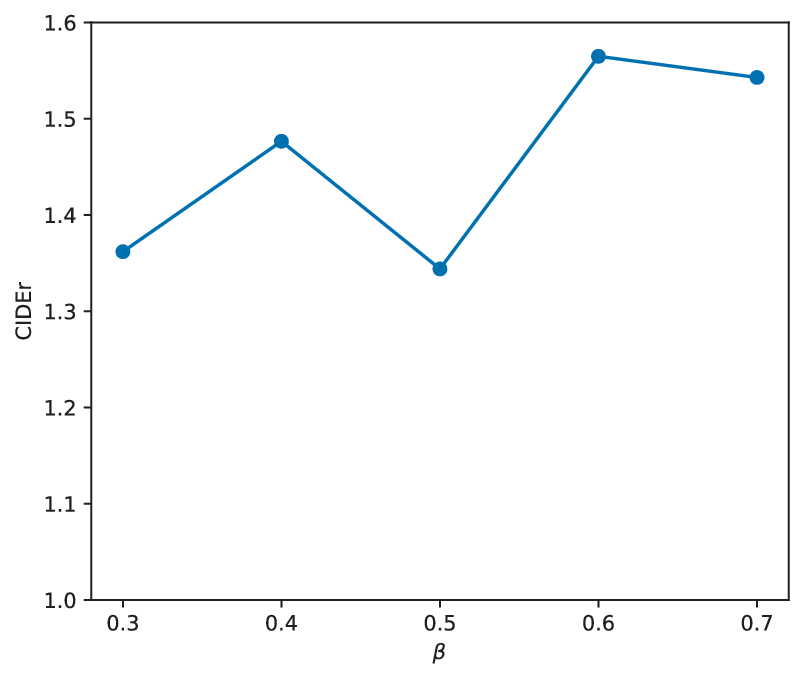}
    \caption{Hyperparameter $\beta$ sensitivity study of \textbf{EIE-light}}
    \label{Hyper-parameter sensitive study of EIE-light}
    \refstepcounter{myfig} 
\end{figure}

In this section, we present a streamlined version of EIE-all, denoted as \textbf{EIE-light}, which relied on expert guidance exclusively during the training phase, excluding the expert soft guidance branch during inference while maintaining commendable performance. To achieve this objective, we propose a random dropout training strategy. Specifically, we assigned an enabling probability $\beta$ to the expert soft guidance. During a random training iteration, we drew a sample from a uniform distribution between 0 and 1, yielding a number $r$. If $r<\beta$, the guidance from the expert classifier was replaced with zero vectors of the same dimension. Otherwise, the guidance was enabled and put into the injection layer. The results of \textbf{EIE-light} are demonstrated in Table \ref{EIE-light comparison}, and the results of the hyperparameter sensitivity study of $\beta$ are illustrated in Figure \ref{Hyper-parameter sensitive study of EIE-light}.

Notably, \textbf{EIE-light} outperformed state-of-the-art methods by a considerable margin, achieving gains of 1.1\%, 2.1\%, 2.9\%, and 3.4\% in BLEU, and 4.3\% and 2.9\% in METEOR and ROUGE$_{\text{L}}$, respectively. Furthermore, there was a notable 0.538 improvement in CIDEr, representing a 52.39\% increase. It is worth mentioning that \textbf{EIE-light} exhibited superior performance compared to \textbf{EIE-mem} and comparable performance with \textbf{EIE-esg}, indicating that even without guidance during inference, the guidance during training significantly enhances the performance of \textbf{EIE-base} and \textbf{EIE-mem}. Additionally, as depicted in Figure \ref{Hyper-parameter sensitive study of EIE-light}, there was minimal fluctuation with changes in $\beta$, maintaining results consistently higher than 1.30, with the optimum performance achieved at $\beta=0.6$.

\begin{table*}[tb]

\renewcommand{\arraystretch}{1.}

\centering

\resizebox{1.\linewidth}{!}{
\begin{tabular}{cccccccc}
\toprule
Method             &BLEU-1          &BLEU-2        &BLEU-3          &BLEU-4          &METEOR           &ROUGE$_\text{L}$           &CIDEr   \\
\hline
\midrule
$\text{MCCFormers}^{\dagger}$         &0.214           &0.190         &0.170           &0.153           &0.319            &0.340           &0.000      \\
$\text{IDCPCL}^{\dagger}$	        &0.614           &0.541         &0.474           &0.414           &0.303             &0.582          &0.703         \\
EKAID              &0.628		    &0.553         &0.491           &0.434           &0.339            &0.577           &1.027           \\
\hline
\hline
EIE-base          &0.574           &0.510         &0.456           &0.407           &0.416            &0.590           &0.968       \\
EIE-mem           &0.614		    &0.547         &0.492           &0.441           &0.405            &0.616           &1.421       \\
EIE-esg           &0.638           &0.573         &0.519           &0.469           &0.399            &0.636           &1.615       \\
\rowcolor{mygray} 
EIE-all               &0.646           &0.583         &0.528           &0.477           &0.401            &0.635           &1.698     \\
\hline
\hline
EIE-light         &0.639           &0.574         &0.520            &0.468            &0.382           &0.606        &1.565  \\
EIE-esg-14         &0.656               &0.591
&0.537        &0.487               &0.408
&0.655        &1.747  \\
EIE-all-14        &0.670               &0.604
&0.549        &0.498               &0.391
&0.639        &1.780 \\

\bottomrule

\end{tabular}
}
\caption{Comparison with State-of-the-Art (SOTA): \textbf{EIE-all} surpasses existing SOTA by a substantial margin across all metrics. Furthermore, each proposed mechanism notably enhances the performance of the base model. The combination of both mechanisms leads to further improvement. $\dagger$ Results were reported from EKAID. $\ddagger$ Results were reproduced using the official codes.}

\label{EIE-light comparison}

\end{table*}

\section{Expert Soft Guidance with More Observations}
As stated in the main paper, we adopted PCAM \cite{ye2020weakly}, which was the top1 solution of CheXpert \cite{smit2020chexbert} classification, as our expert classifier. However, PCAM can only produce the predictions of the 5 most common diseases contained in the CheXpert test set, resulting in suboptimal guidance. Therefore, it is intuitive that the performance would be further improved if the expert classifier could provide guidance on more observations. A recent work \cite{pellegrini2023radialog} is capable of offering the classification results for all 14 observations of CheXpert, denoted as \textbf{Expert-14}, which was then incorporated into our model and was assessed accordingly. When the expert classifier of \textbf{EIE-esg} was substituted by \textbf{Expert-14}, we denote it as \textbf{EIE-esg-14}. When the classifier of \textbf{EIE-all} was replaced with \textbf{Expert-14}, it is denoted as \textbf{EIE-all-14}.

All the results are demonstrated in Table \ref{EIE-light comparison}. We can see that model performance improved further when more guidance was available. Specifically, \textbf{EIE-esg-14} brought more performance gain on all metrics than \textbf{EIE-esg}: 1.8\%, 0.9\%, 1.9\% and 0.132 on BLEU, METEOR, ROUGE$_\text{L}$ and CIDEr, respectively. Together with the guidance from all 14 observations and MEM, \textbf{EIE-all-14} consistently achieved further improvements in performance. Compared with \textbf{EIE-all}, \textbf{EIE-all-14} gained 2.4\%, 2.1\%, 2.1\% and 2.1\% in BLEU, 0.4\% and 0.082 in ROUGE$_\text{L}$ and CIDEr respectively. These results suggest that more expert guidance could further boost the overall performance of our proposed model. It is worth noting that although PCAM can only predict the presence of the most common abnormalities on chest X-ray, it can significantly improve the overall performance. This could be due to the fact that PCAM can not only offer the presence probability of these abnormalities, but also the presence probability of other related abnormalities.

\section{Entity Selection}
To enrich expert insight into abnormality prediction and complement \textbf{EIE-esg}, we have selected relevant terms associated with diseases from CheXpert, namely atelectasis, edema, pneumothorax, cardiomegaly, consolidation, cardiac silhouette, fracture, lung opacity, pleural effusion, and pneumonia.

\section{Metrics}
Following \cite{tanida2023interactive}, we have incorporated clinical efficacy metrics, Acc5 and Acc14, to showcase the effectiveness of our model. Specifically, Acc5 represents the micro-averaged accuracy across the 5 most common observations: atelectasis, cardiomegaly, consolidation, edema, and pleural effusion. Acc14 is the example-based averaged accuracy across all 14 observations in CheXpert, encompassing pneumonia, fracture, consolidation, enlarged cardiomediastinum, no finding, pleural other, cardiomegaly, pneumothorax, atelectasis, support devices, edema, pleural effusion, lung lesion and lung opacity. We employed the code from \cite{tanida2023interactive} to calculate these clinical efficacy metrics, utilizing CheXbert \cite{smit2020chexbert} labeler to determine the presence of the most common observations. It is noteworthy that, following \cite{tanida2023interactive}, instances labeled as \textit{negative, uncertain, or no mention} were all treated as negative results.

\bibliographystyle{splncs04}
\bibliography{icme2023template}